\documentclass[10pt,letterpaper]{article}
\usepackage{opex3}
\usepackage{hyperref}
\usepackage[space]{cite}
\usepackage{amsmath}

\begin{document}

\title{Wideband laser locking to an atomic reference with modulation transfer spectroscopy}

\author{V. Negnevitsky$^*$ and L. D. Turner}
\address{School of Physics, Monash University, VIC 3800, Australia\\
$^*$ Present address: Institute for Quantum Electronics, ETH Z\"{u}rich, 8093 Z\"{u}rich, Switzerland.}
\email{nvlad@phys.ethz.ch}

\begin{abstract*}
We demonstrate that conventional modulated spectroscopy apparatus, used for  laser frequency stabilization in many atomic physics laboratories, can be enhanced to provide a wideband lock delivering deep suppression of frequency noise across the acoustic range.
Using an acousto-optic modulator driven with an agile oscillator, we show that wideband frequency modulation of the pump laser in modulation transfer spectroscopy produces the unique single lock-point spectrum previously demonstrated with electro-optic phase modulation.
We achieve a laser lock with 100\,kHz feedback bandwidth, limited by our laser control electronics. This bandwidth is sufficient to reduce frequency noise by 30\,dB across the acoustic range and narrows the imputed linewidth by a factor of five.
\end{abstract*}
%

\section{Introduction}
Closed-loop laser frequency stabilization --- `laser locking' --- is required for laser cooling, precision spectroscopy and atomic clocks.
These experiments demand laser light frequency-locked to an atomic transition to an absolute accuracy of better than 1\,MHz.
Such accuracy may be achieved by stabilizing the laser to a sub-Doppler spectroscopic feature of an atomic reference, with several locking schemes in common use.
Schemes where the laser is frequency-modulated~\cite{hall_optical_1981,bjorklund_frequency_1983,bell_laser_2007} are often preferred to baseband schemes~\cite{pearman_polarization_2002,robins_interferometric_2002} as they operate at frequencies above most laser technical noise and are less sensitive to alignment drift, temperature variations and stray light.
The most common modulation schemes are frequency modulation spectroscopy~(FMS)~\cite{hall_optical_1981} and  modulation transfer spectroscopy~(MTS)~\cite{camy_heterodyne_1982}.
We show that modulation transfer spectroscopy is particularly suited to wideband locking, providing strong acoustic noise rejection and appreciable linewidth narrowing, advantages typically associated with systems using an optical cavity as a reference.
The system retains the environmental immunity, robustness and absolute accuracy of a direct atomic lock.
Using only the optics of a commonly used acousto-optic modulator (AOM) lock setup, we show that enhanced electronics delivers feedback bandwidths over 100\,kHz, suppressing frequency noise across the acoustic range by 30\,dB.
Furthermore, we demonstrate that as the modulation frequency is increased into the megahertz range, the error signal evolves from the familiar cluster of overlapping dispersion curves into a characteristic modulation transfer spectrum with an unambiguous lock point located only at the closed atomic transition.

There are several reasons to prefer MTS over FMS for locking purposes.
In both schemes the probe is coherently demodulated to obtain the error signal.
The FMS lineshape arises due to the direct probing of the vapor absorption and dispersion by the probe beam~\cite{bjorklund_frequency_1983}, while MTS relies on the frequency-dependent \emph{transfer} of modulation from the pump to the probe beam~\cite{shirley_modulation_1982}.
The modulation frequency $f_m$ of an MTS setup is limited roughly to the natural linewidth, above which the error signal lock point develops a `kink' \cite{mccarron_modulation_2008}.
FMS avoids this limitation, offering higher modulation frequencies and potentially higher signal-noise ratio (SNR).
However, as it is still a linear process, FMS is encumbered with an undesirable Doppler background, and the amplitudes of its sub-Doppler features match those of the saturated absorption spectrum without emphasizing the closed transitions; which can lead to the lock point `hopping' from one transition to another, among other pathologies.
FMS is also sensitive to any dispersive element, including parasitic etalons, which may change the lock point by adding offsets to the demodulated error signal.
In contrast, MTS produces symmetric sub-Doppler features free of Doppler background that are strong at closed transitions, such as $^{87}$Rb $F=2 \rightarrow F'=3$. Undesirable crossovers and other transitions are suppressed (given the correct selection of MTS parameters) and parasitic etalons do not contribute offsets. 

Many laboratories employ a low-frequency variant of MTS, frequency modulating the pump beam with an acousto-optical modulator at relatively low modulation frequencies of order 100\,kHz.
This spectroscopy mode yields spectra without the Doppler background and free of etalon offsets, but the spectra otherwise resemble FM spectra showing all transitions and crossovers, and the usable servo bandwidth is less than 10\,kHz.
Recent work has revived interest in MTS-based laser locking, in particular due to the clean spectra and wide control bandwidths made possible by MTS at high modulation frequencies~\cite{mccarron_modulation_2008,eble_optimization_2007}. 
These workers used resonant electro-optical modulators (EOMs) to phase modulate the pump beam at high modulation frequencies above 1\,MHz.
While AOMs have been used in both FMS and MTS systems~\cite{camy_heterodyne_1982,du_burck_narrow-band_2003} to provide improved optical isolation and intensity noise suppression~\cite{xiang-hui_ultra-stable_2009,du_burck_narrow_2009}, to our knowledge there has been no detailed presentation of wideband AOM-based MTS locking. 
The key result of this paper is that a conventional modulated spectrometer using an acousto-optic modulator can deliver wideband modulation and coherent modulation transfer spectra using little, if any, additional optics.
This AOM-based MTS spectrometer inherits the unambiguous spectrum, Doppler-free background, immunity to etalon interference effects and wide control bandwidth of EOM-based spectrometers, and requires only changes to the electronics operating a conventional AOM-based pump-probe spectrometer.
We discuss the design and performance of such a system, and the various considerations in using it for robust laser locking.
\section{Experimental setup}
\begin{figure}[tbp]\centering
	\includegraphics[]{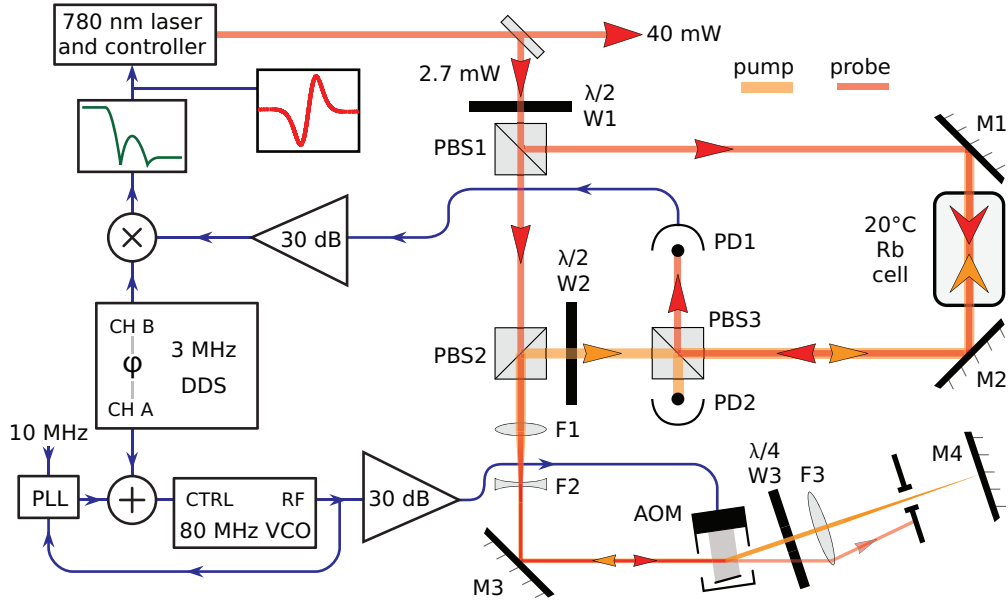}
	\caption{Electro-optical schematic of our acousto-optic MTS system. Frequency modulation is imparted to the pump beam via a double pass of the acousto-optic modulator (AOM). Modulation is transferred from the pump to the probe within the rubidium vapor cell. The beat of the probe frequency components at photodetector PD1 (12\,MHz bandwidth) is demodulated to produce an error signal. F1, F2, F3 = +100, -35, +100 mm.}
	\label{fig:mts_optical_layout}
\end{figure}
In our spectrometer (Fig.~\ref{fig:mts_optical_layout}) an AOM~\cite{equipment_list_optical} shifts the frequency of the pump beam by 160\,MHz and imposes a fast frequency modulation.
This results in a lock point 80\,MHz below the rubidium D$_2$ $F=2 \rightarrow F'=3$ cycling transition, which is optimal for subsequent acousto-optical generation of several near-resonant beams.
The AOM shift has the addition benefit of providing optical isolation between the diode laser and the spectrometer.
We use an external-cavity diode laser~\cite{hawthorn_littrow_2001}, with a commercial laser controller \cite{equipment_list_optical} for temperature stabilization and control of the diode injection current and laser cavity length.
The laser produces a 40\,mW elliptical beam with a large horizontal eccentricity of 0.8 and a D4$\sigma$ (second moment) diameter of 1.4\,mm~\cite{d4sigma_definition}.
We direct 2.7\,mW to the spectrometer, of which 250\,\textmu W is used for the probe beam.
The pump beam passes through a telescope (F1, F2) to reduce its diameter to 500\,\textmu m with a vertical eccentricity of 0.8, and is double-passed through the AOM using a `cat-eye' retroreflector made up of F3 and M4~\cite{donley_double-pass_2005}.
The double pass reduces unwanted wavevector modulation and amplitude modulation compared to single-pass arrangements such as those described in Refs.~\cite{jaatinen_iodine_2003} and~\cite{zhang_characteristics_2003}.
Narrowing the beam improves AOM modulation bandwidth, and collimation minimizes the range of Bragg angles, reducing spatial inhomogeneity in the modulation of the pump beam.
The frequency-modulated pump beam incident on the room-temperature rubidium vapor cell is 900\,\textmu W with the same vertical eccentricity, and a diameter of 2.4\,mm.

The 80\,MHz rf drive for the AOM is generated by an agile voltage-controlled oscillator (VCO), locked to a 10\,MHz reference using a phase-locked loop~(PLL) evaluation board  \cite{equipment_list_electronic}.
A direct digital synthesizer (DDS) board generates the modulating signal.
An $f_m=3$\,MHz modulation was used for the laser lock results presented in Section~5, added to the PLL control signal on a bias tee \cite{equipment_list_electronic}.
The PLL loop bandwidth is 5\,kHz, being deliberately low to avoid the PLL feedback interfering with the modulation.
The system is easily adaptable to other AOM center frequencies, provided a VCO is available with sufficient modulation bandwidth.

The pump and probe counter-propagate through the vapor cell.
Two processes --- modulated optical pumping and four-wave mixing --- compete to transfer modulation from the pump to the probe~\cite{shirley_modulation_1982}; these are discussed in more detail in the next section.
The result of the modulation transfer is the generation of sidebands on the probe beam, offset from the optical carrier by the modulation frequency.
Probe sidebands beat with the carrier at a photodetector \cite{photodetector_details}, producing a 3\,MHz signal at approximately -50\,dBm. 
After amplification by 30\,dB the probe signal is demodulated by mixing with the modulating signal, phase-shifted by demodulation phase $\phi$, on a phase detector \cite{equipment_list_electronic}.
Accurate phase shifts were set digitally using a second channel of the DDS.
After the  mixer, a fifth-order modified Chebyshev low-pass filter removes up-converted noise and image components, yielding a clean wideband error signal.
The Chebyshev filter provides a rapid transition from passband to stopband, at the expense of non-monotonic response in the stopband. 
We exploit this non-monotonicity by placing two notches at 3 and 6\,MHz ($f_m$ and $2f_m$).
The filter gain remains nearly flat until 2\,MHz, providing a much wider passband than simpler filter designs while ensuring adequate attenuation in the key regions of the stopband.
Finally the error signal is fed back to the laser controller, where the external-cavity diode current (proportional-integral) gain and piezo (double integrator) gain are individually adjustable.
\section{Modulation transfer lineshapes}
\begin{figure}[btp]\centering
	\includegraphics[]{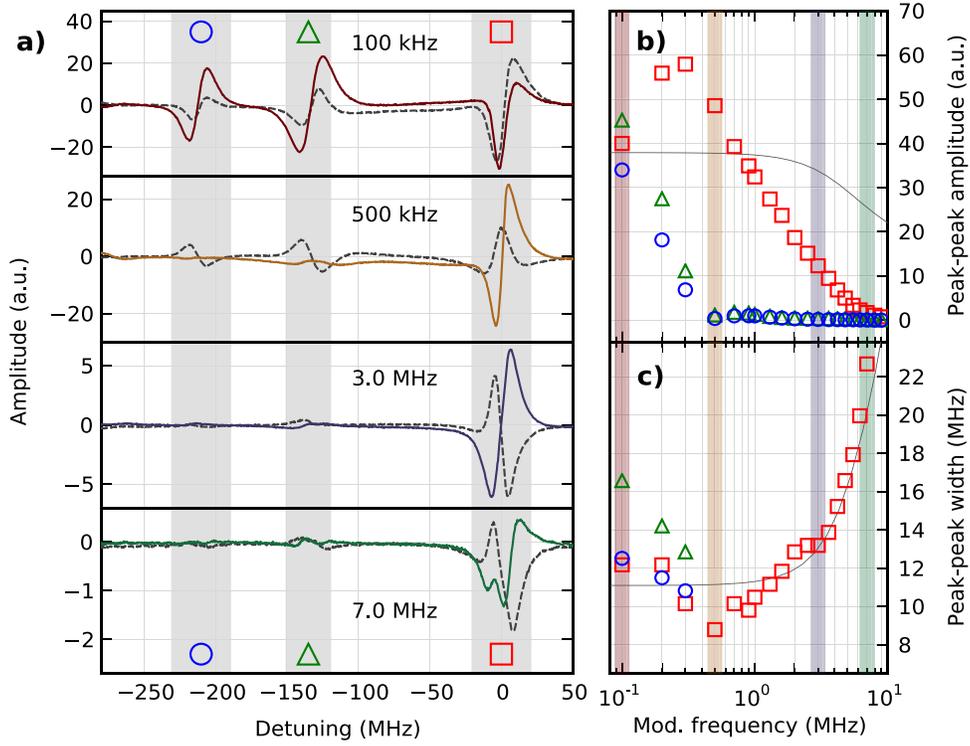}
	\caption{Variation of MTS spectra with modulation frequency: \textbf{a)} solid (dashed) spectra represent 0\textdegree~(90\textdegree) components, with \textbf{b)} amplitudes and \textbf{c)} trough-to-peak widths of each spectral feature at 0\textdegree. Solid lines in \textbf{b)} and \textbf{c)} represent the predictions of Eq. 1 in \protect{\cite{jaatinen_theoretical_1995}} using a linewidth value of 9.5\,MHz; the line in \textbf{b)} is scaled to the experimental amplitudes using a least-squares fit. Widths are omitted where signal amplitude is too small for their reliable determination. [blue circle] ($F=2 \rightarrow F'= \{ 1,2 \}$) and [green triangle] ($F=2 \rightarrow F'= \{ 1,3 \}$) are crossover resonances. [red square] ($F=2 \rightarrow F'=3)$ is the closed `cooling' transition.}
	\label{fig:mts_spectra}
\end{figure}

Our use of acousto-optic modulation makes it much easier to vary the modulation frequency $f_m$ than with a resonant EOM, helping us find the modulating frequency giving the optimal lock and a clean MTS spectrum.
We studied the MTS lineshape as $f_m$ was varied from 100\,kHz to 8\,MHz, limited by the mixer below 100\,kHz and the AOM modulation bandwidth above 8\,MHz.
This provided novel information on MTS in rubidium below 1\,MHz, a region where it is difficult to construct high-Q resonant electro-optic modulators~\cite{mudarikwa_sub-doppler_2012}, bridging the gap between AOM-based schemes with $f_m < 200$\,kHz~\cite{zhang_characteristics_2003,zhou_observation_2010} and EOM-based schemes with $f_m$ above several megahertz~\cite{mccarron_modulation_2008}.
We maintained a \emph{constant frequency deviation} of 2.4\,MHz (i.e. maximum deviation of the rf from 80\,MHz, limited by our VCO); larger deviations produce monotonically larger locking features, but the spectra are not significantly altered otherwise.
Recent studies of MTS have used a \emph{constant modulation index}~\cite{mccarron_modulation_2008, mudarikwa_sub-doppler_2012}, which precludes direct comparison of the results due to the nonlinear dependence of feature size upon frequency deviation.
Demodulated spectra presented were filtered through a 20\,kHz Bessel low-pass filter, rather than the wideband loop filter used for locking; Fig.~\ref{fig:mts_spectra}(a) displays several example spectra.

In MTS, frequency modulation on the pump beam is converted to amplitude modulation~(AM) on the probe beam through the mutual interaction of the beams with the atomic reference.
The frequency-dependent AM phase is recovered via demodulation to produce a symmetric error signal.
Residual amplitude modulation~(RAM) of the pump beam distorts the error signal, and occurs due to the asymmetric frequency responses of both the AOM and VCO.
The distortion is an evenly symmetric sum of Lorentzians centered around the transition frequency~\cite{jaatinen_compensating_2008}, shifting the zero crossing of the error signal away from resonance.
RAM may be minimized at a chosen modulation frequency by careful optical alignment, as discussed in the next section. 
For the data presented in Fig.~\ref{fig:mts_spectra}, RAM was minimized once at $f_m = $3\,MHz, and so spectra are increasingly distorted by RAM for higher $f_m$.
The spectra for lower $f_m$ are much less affected by RAM, and the distortions present especially in the cycling transition are discussed later in this section.

Contributions to MTS spectra may be broadly classified as incoherent (modulated hole burning) or coherent (four-wave mixing). 
The most familiar process is hole burning, whereby the pump beam creates a Bennett hole in the ground state population, and the probe beam measures reduced absorption when both probe and pump beam are resonant with atoms of the same velocity class.
This process occurs with unmodulated probe beams, and is commonly known as `saturated absorption'; it is well-known that in the alkalis the reduction of absorption on open transitions and crossovers is in fact largely due to the incoherent process of `velocity selective optical pumping' into the other hyperfine ground state~\cite{smith_role_2004}.

At low modulation frequencies, the pump sweeps across the ground state velocity distribution, modulating the location of the Bennett hole on the distribution.
The absorption of the velocity class resonant with the probe is thus modulated, causing in-phase amplitude modulation on the probe beam.
This process is clearest in the $f_m=100$\,kHz trace uppermost in Fig.~\ref{fig:mts_spectra}. 
As a simple sub-Doppler modulation spectroscopy, the spectrum resembles the derivative of the unmodulated absorption spectrum and shows strong signals at crossover resonances (leftmost two features) and closed transitions (rightmost feature).
It thus resembles an FMS spectrum, though without the Doppler background.
At much higher modulation frequencies, the period of the frequency modulation becomes short compared to the mean transit time of an atom across the probe beam.
In this limit there will be atoms still transiting the beam which have been optically pumped by the pump beam at any given optical frequency within its deviation range, and so there will be vanishing oscillation of the probe absorption at the modulation frequency.
For our beam diameters and a room-temperature thermal distribution, the mean transit time is equal to the modulation period at approximately $f_m=150$\,kHz. 
Figure~\ref{fig:mts_spectra}(b) shows the heights of the crossover resonances falling sharply~\cite{amp_width_calculation} and becoming unmeasurable by $f_m=400$\,kHz, consistent with this simple model of modulated optical pumping.

The coherent component of a modulation transfer spectrum has been attributed to four-wave mixing, whereby the pump carrier, probe carrier and a pump sideband interact via the $\chi^3$ susceptibility of the near-resonant vapor to create a new probe sideband~\cite{shirley_modulation_1982}.
This process is inoperative at DC, is largely (if not entirely~\cite{noh_modulation_2011}) suppressed at crossover resonances, and is only efficient at closed transitions.
The incoherent process weakens as the modulation frequency is increased, revealing the coherent single-feature spectra at frequencies above 1\,MHz.
Below 1\,MHz modulated optical pumping is much more significant than four-wave mixing for the crossover lineshapes, however both processes contribute significantly to the closed transition.
For example, at 100\,kHz the crossovers are relatively undistorted, while the closed transition shows strong asymmetry.
This distortion appears to be independent of RAM effects, which should affect all spectral features equally.
We cannot conclusively pinpoint its cause, although it appears to be due to the interplay between the two MTS processes.

The spectra (Fig.~\ref{fig:mts_spectra}(a)) and peak-to-peak heights (Fig.~\ref{fig:mts_spectra}(b)) show that while the absolute height of the closed transition feature falls relatively slowly (approximately as the negative logarithm) with modulation frequency, the closed transition completely dominates the spectrum for $f_m>500$\,kHz, providing an unambiguous lock point.
The roll-off of the closed transition height in Fig. \ref{fig:mts_spectra}(b) is steeper than predicted by Eq.~(1) in Ref.~\cite{jaatinen_theoretical_1995} using a simple model of the atomic four-wave mixing process.
We attribute this discrepancy to the limited modulation bandwidth of the AOM, discussed in the following section.

The peak-trough widths of the spectral features are shown in Fig.~\ref{fig:mts_spectra}.
We fit these to the same simple model, using the effective linewidth as the free parameter.
Agreement is best with an effective linewidth of 9.5\,MHz (1.6 times the natural linewidth).
As noted elsewhere, the four-wave mixing process cannot be modeled as a simple two-level system~\cite{mccarron_modulation_2008}.
This is clearly illustrated by the fact that a two-level model predicts a power-broadened linewidth of around 20\,MHz.

We found that $f_m=3$\,MHz was optimal for locking purposes, providing spectra with weak crossovers, a single clear lock point at the closed transition and a high SNR.
Frequencies well below 3\,MHz were insufficient to escape $1/f$ environmental and laser noise, and the limited modulation bandwidth of the AOM reduced the signal amplitude above 3\,MHz.
Modulating at 3\,MHz should give a usable servo bandwidth of order 1\,MHz for a laser controller able to respond sufficiently fast. 
\section{Optimizing the spectrometer}
Our MTS-based lock employs a high modulation frequency of 3\,MHz, delivering a usable feedback bandwidth of several hundred kHz. 
Working at such modulation frequencies places more critical demands on AOM alignment than those of commonly-used narrowband modulation at 100\,kHz or below.
We briefly summarize techniques we found useful in achieving a symmetrical error signal at these high frequencies.

AOMs diffract efficiently when the Bragg criterion is satisfied, connecting angles of incident, diffracted and radiofrequency waves. 
Conventionally this match is best made with the widest possible laser beam, limited by the `acoustic aperture' of the AOM.
Such wide beams present no problems at conventional low modulation frequencies: at $f_m=100$\,kHz the acoustic wave travels several cm over a modulation cycle, and the laser beam illuminates a moving grating of spatial homogeneous period.
At our high modulation frequency of $f_m=3$\,MHz in our AOM crystal (TeO$_2$, $v_\text{sound}=4.2$\,mm/\textmu s), the wave propagates only 1.4\,mm during a cycle, so that the a 1\,mm wide diffracted beam contains almost the full cycle of frequency deviations across its transverse profile.
This `modulation bandwidth' limit reduces contrast of the MTS error signal, and we reduce the incident beam diameter to 500\,\textmu m to ameliorate this effect.
Focusing the pump beam through the AOM tends to produce too narrow a waist for reasonable focal lengths, and instead we found that forming a collimated beam of diameter 500\,\textmu m with a telescope (as shown in Fig.~\ref{fig:mts_optical_layout}) yields the best compromise of sufficient diffraction efficiency with only slight compression of the MTS error signal due to the acoustic transit-time effect.

Operation of the cat-eye AOM double pass has been thoroughly described elsewhere~\cite{donley_double-pass_2005}. 
A convenient focal length for cat-eye lens F3 was 100\,mm, with F3 placed 100\,mm after the AOM and the retroreflection mirror M4 a further 100\,mm after the lens. 
In a variation on the layout of Ref.~\cite{donley_double-pass_2005}, the lens axis and mirror normal are co-linear with the diffracted beam axis.
This alignment is easily made using an optical cage bench attached to a kinematic mount.
During this alignment, attention was paid to 
(i) maximizing double-pass efficiency and 
(ii) minimizing the deflection and distortion of the double-passed beam induced by varying the rf carrier frequency (by several MHz).

After minimizing the angular deviation of the double-passed beam, we next measured the residual amplitude (intensity) modulation using a photodiode~(PD2 in Fig.~\ref{fig:mts_optical_layout}). 
On a DC-coupled oscilloscope, residual amplitude modulation (RAM) appears as a small $f_m = 3$\,MHz signal, superposed on a large DC offset corresponding to the diffracted beam intensity. 
Slight rotations of the cage bench and M3, and small adjustments of the carrier radiofrequency, minimize the RAM, while aiming to preserve the diffracted beam power.
When the optics and electronics were optimally adjusted the RAM minimum coincided with the diffracted intensity (DC) maximum.
The constant time delay due to the AOM, caused by the travel time of the acoustic wave from the piezoelectric transducer to the point of interaction with the light, can be accurately measured by amplitude modulating the AOM and demodulating the signal from PD2.

The final optical step is the careful superposition of the modulated pump beam with the probe beam in the vapor cell.
This alignment was best made using the MTS signal as a diagnostic, adjusting beam overlap until a symmetric error signal was obtained.
The signal was observed at several settings of the demodulation phase.
We found that imperfect suppression of amplitude modulation in the AOM could be largely compensated with beam overlap~\cite{jaatinen_compensating_2008}. 

Electronic adjustments complete the optimization.
The demodulation phase was chosen to maximize the error signal amplitude, and the lineshape was balanced using the above techniques for this particular phase.
Finally, we note that small frequency shifts in the lock point of up to several megahertz may be realized with great precision and stability by altering the AOM carrier frequency.
We anticipate that such `frequency shimming' will be useful to calibrate the MTS lock point to a more precise atomic reference, such as absorption measurements of cold atoms in regions of minimal stray magnetic field.
\section{Lock performance}
The major aims of this work were to produce a stable, reliable lock, with strong frequency noise suppression and significant linewidth reduction.
Reliability was verified empirically over a period of months: typically the laser remains locked for many hours, and often days, at a time, with loss of lock caused most often by laser mode hops or temperature fluctuations exceeding the controller range.
The lock is difficult to dislodge without mechanically disturbing the laser itself, thus no special care needs to be taken when working near the system.

\begin{figure}[tbp]\centering
	\includegraphics[]{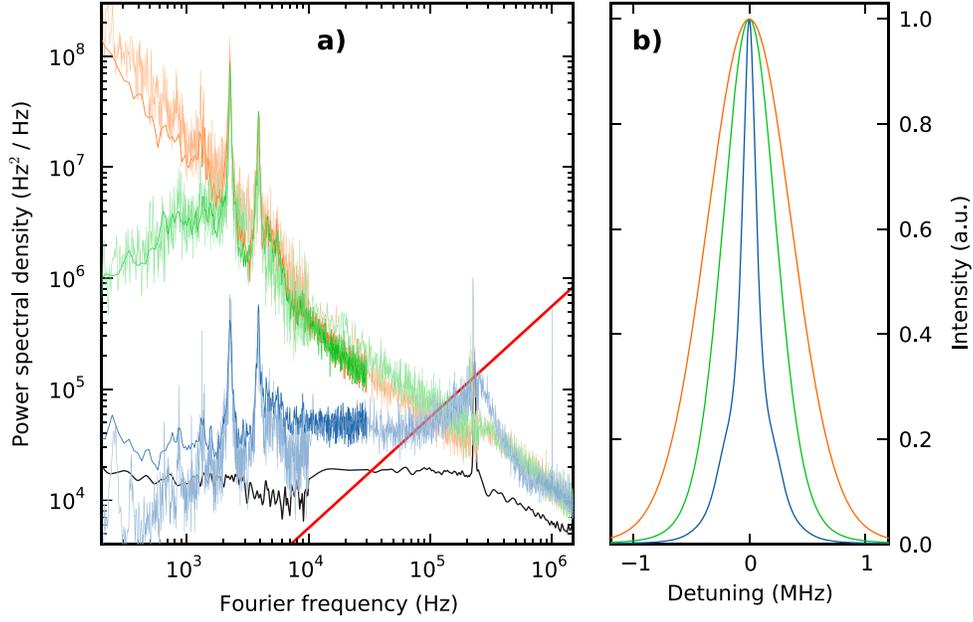}
	\caption{Laser stabilization properties. \textbf{a)} Composite $S_{\delta\nu}(f)$, showing both EagleEye and MTS error signal frequency noise, for (top to bottom) piezo feedback with minimal gain, piezo and current feedback with minimal gains, piezo and current with optimal gains. Dark (light) traces represent EagleEye (MTS error signal) data. Below 30\,kHz both are displayed, while above 30\,kHz only MTS error signal data are shown. The $\beta$-separation line (red) and off-resonant MTS error signal noise floor (black) are shown (EagleEye photodetector noise floor is below this by 30\,dB). \textbf{b)} (outer to inner) corresponding inferred lineshapes, calculated from spectra in (a) using $S_{\delta\nu}(f)$ from EagleEye data below 30\,kHz, and MTS error signal data from 30\,kHz to 1.5\,MHz.}
	\label{fig:freq_noise}
\end{figure}

To quantify noise suppression and linewidth, we measured laser frequency noise at Fourier frequencies up to 30\,kHz using an optical spectrum analyzer (Sirah EagleEye) based on a high-finesse cavity. 
The EagleEye signal processor servos the cavity length to maintain the laser frequency half-way up an Airy fringe of the cavity transmission spectrum, and records time series of the transmission.
Welch's average periodogram method~\cite{press_numerical_1992} was used to obtain the power spectral density~(PSD) $S_{\delta\nu}(f)$ of the laser frequency noise, shown in Fig.~\ref{fig:freq_noise}(a).
From 30\,kHz to 1.5\,MHz the pre-filter error signal spectrum was measured electronically, and calibrated to the EagleEye data by measuring the response of both to modulation of the controller input.

The frequency noise spectrum is more fundamental than the Allan variance and generally a more helpful diagnostic~\cite{turner_frequency_2002}, and was used to optimize the feedback gains.
Figure \ref{fig:freq_noise} shows the noise suppression and imputed linewidth narrowing of the MTS system.
A low-gain piezoelectric-only laser lock was used to estimate the free-running laser frequency noise, holding the laser frequency steady during measurements (top trace in Fig.~\ref{fig:freq_noise}(a)).
To leverage the wideband design of the spectrometer, we added injection current feedback (second trace), which strongly attenuated noise below 2\,kHz but failed to suppress mechanical resonances at 2 and 4\,kHz.
Piezo and current gains were then optimized for maximum noise suppression, raising the bandwidth up to 100\,kHz with over 30\,dB suppression across acoustic frequencies (third trace).
The peaks in the noise spectrum below 5\,kHz are due to mechanical resonances in the laser grating mount.

A common figure of merit for laser frequency stability is full-width at half-maximum (FWHM) linewidth, typically measured for sub-megahertz lines using a beatnote between two similar laser systems.
Lacking two lasers, we instead inferred the optical spectrum from the PSD by determining the autocorrelation function, then applying the Wiener-Khinchin theorem to reconstruct the lineshape~\cite{elliott_extracavity_1982}.
The effect on the FWHM linewidth of noise at Fourier frequency $f$ hinges primarily on the strength of the noise $S_{\delta\nu}(f)$ relative to the line $8 \ln(2) f / \pi^2$, which separates the PSD into regions of high and low modulation index~\cite{di_domenico_simple_2010}.
Noise above this `$\beta$-separation' line (i.e. at frequencies where $S_{\delta\nu}(f) > 8 \ln(2) f / \pi^2$) causes a relatively high frequency deviation compared to its Fourier frequency, and contributes significantly to broadening the line at its shoulders.
The further $S_{\delta\nu}(f)$ is above the line, the more the FWHM linewidth is increased.
The converse applies to noise `below the line', which affects only the wings of the lineshape, and its suppression is unimportant for reducing the FWHM linewidth.

The control bandwidth of our system was sufficient to suppress noise across the entire frequency range where it exceeded the $\beta$-separation line.
Figure \ref{fig:freq_noise}(a) shows the frequency noise suppressed to almost the system noise floor from near DC up to  60\,kHz.
As with all stable controllers to which the Bode sensitivity integral applies, a distinct `servo bump' appeared above the 0\,dB point (100\,kHz) as a necessary consequence of noise suppression at lower frequencies.
This bump is visible as a broadening of the base in Fig.~\ref{fig:freq_noise}(b), however because it falls largely below the $\beta$-separation line the FWHM linewidth is not significantly increased by its presence.

For controller gains higher than those used for the `optimal' trace in Fig.~\ref{fig:freq_noise}, the noise in the error signal fell below the EagleEye frequency noise floor, as photodetector noise and laser intensity noise began to dominate frequency noise in the control loop.
In this limit the error signal was an unreliable measure of linewidth.
To further reduce the linewidth our system would require significantly higher SNR of the error signal; this could be achieved by increasing the power diverted to the spectrometer, increasing beam diameters, reducing noise by using intensity noise cancellation~\cite{du_burck_narrow_2009} and/or heating the vapor cell.

Laser systems are often compared using the root mean square (RMS) of their frequency noise to obtain a `linewidth', especially where a frequency discriminator is used~\cite{hadrich_narrow_2008,thompson_narrow_2012}.
As shown in Tab. \ref{tab:locking_linewidths}, RMS linewidths poorly reflect the valuable FWHM linewidth reduction visible in Fig.~\ref{fig:freq_noise}(b).
RMS figures underestimate the linewidth when $1/f$ noise dominates and overestimate it when it is suppressed, due to the dominance of the servo bump.
\begin{table}[tbp]\centering
  \begin{tabular}{c c c c c}
    Lock actuators (gain) & FWHM & RMS\\
    \hline
    Piezo (minimal) & 867 & 389 \\
    Piezo + current (minimal) & 547 & 275 \\
    Piezo + current (optimal) & 165 & 216
  \end{tabular}
\caption{FWHM and RMS linewidths under the locking circumstances of \protect{Fig.~\ref{fig:freq_noise}}. All quantities in kilohertz.}

\label{tab:locking_linewidths}
\end{table}

Our system has thus reduced the imputed laser linewidth by a factor of 5, limited by the controller bandwidth of 100\,kHz and the noise floor of the system.
We hope to increase this by extending the control bandwidth to at least several hundred kilohertz; the wideband MTS electronics should support this without modification. 
The PSD and laser lineshape are both useful for designing and characterizing such an extension.
\section{Conclusions}
In summary, this paper has outlined our approach for designing, characterizing and optimizing an AOM-based laser lock using modulation transfer spectroscopy.
We have combined conventional, inexpensive AOM-based optics with a wideband electronic modulation and demodulation scheme, and shown that the error signal is of sufficient quality to reduce the inferred linewidth of a typical external-cavity diode laser by a factor of five. 
We have also demonstrated that to obtain the full range of benefits an MTS spectrometer provides, modulation frequencies above 500\,kHz must be used.
Heuristics for designing and optimizing such a system have been presented, which should help improve performance for adopters of this system.
The laser frequency PSD has been used to quantify lock performance, and numerical calculation and analysis of the laser lineshape will inform future improvements to the controller.
\section*{Acknowledgments}
This work was supported by a Monash University Faculty of Science ECR Grant, the J.\,L.~William Fund and the Australian Research Council (DP1094399). R.\,P.~Anderson, M.~Jasperse, A.\,A.~Wood and L.\,M.~Bennie assisted with the construction and optimization of the laser system. We are obliged to R.\,E.~Scholten for loan of the EagleEye optical spectrum analyzer. The manuscript benefited greatly from a thorough review by R.\,P.~Anderson and R.\,E.~Scholten, for which we are very grateful. We thank A.~Slavec for implementing the Chebyshev filter and A.~Benci for electronics support.

\begin{thebibliography}{10}
\bibitem{d4sigma_definition}
Four times the geometric mean of the horizontal and vertical standard deviations of the beam intensity distribution. D4$\sigma$ diameter is equivalent to $1/e^2$ diameter for Gaussian beams, and is less affected by noise for non-Gaussian beams such as those used in the spectrometer.
\bibitem{amp_width_calculation}
The peak-to-peak height and width of each spectral feature were obtained by numerically locating the outermost pair of stationary points within the expected frequency range of the spectral feature and above a threshold amplitude, then calculating the frequency and amplitude difference between them.
This method is immune to RAM, which causes a distortion with even symmetry around the transition frequency.
The MTS features are odd-symmetric around this frequency, thus RAM distortion shifts both stationary points by a common distance in frequency and amplitude.
Other stationary points, such as the central trough in the 7\,MHz closed transition feature (due to high levels of RAM), are ignored by the algorithm.
\end{thebibliography}
\end{document}